# Repetitive TMS-based Identification of Methamphetamine-Dependent Individuals Using EEG Spectra

Ziyi Zeng, Yun-Hsuan Chen, Xurong Gao, Wenyao Zheng, Hemmings Wu, Zhoule Zhu, Jie Yang, *Senior Member, IEEE*, Chengkai Wang, Lihua Zhong, Weiwei Cheng, and Mohamad Sawan, *Life Fellow, IEEE*

*Abstract*—The impact of repetitive transcranial magnetic stimulation (rTMS) on methamphetamine (METH) users' craving levels is often assessed using questionnaires. This study explores the feasibility of using neural signals to obtain more objective results. EEG signals recorded from 20 METH-addicted participants Before and After rTMS (MBT and MAT) and from 20 healthy participants (HC) are analyzed. In each EEG paradigm, participants are shown 15 METH-related and 15 neutral pictures randomly, and the relative band power (RBP) of each EEG sub-band frequency is derived. The average RBP across all 31 channels, as well as individual brain regions, is analyzed. Statistically, MAT's alpha, beta, and gamma RBPs are more like those of HC compared to MBT, as indicated by the power topographies. Utilizing a random forest (RF), the gamma RBP is identified as the optimal frequency band for distinguishing between MBT and HC with a 90% accuracy. The performance of classifying MAT versus HC is lower than that of MBT versus HC, suggesting that the efficacy of rTMS can be validated using RF with gamma RBP. Furthermore, the gamma RBP recorded by the TP10 and CP2 channels dominates the classification task of MBT versus HC when receiving METH-related image cues. The gamma RBP during exposure to METH-related cues can serve as a biomarker for distinguishing between MBT and HC and for evaluating the effectiveness of rTMS. Therefore, real-time monitoring of gamma RBP variations holds promise as a parameter for implementing a customized closed-loop neuromodulation system for treating METH addiction.

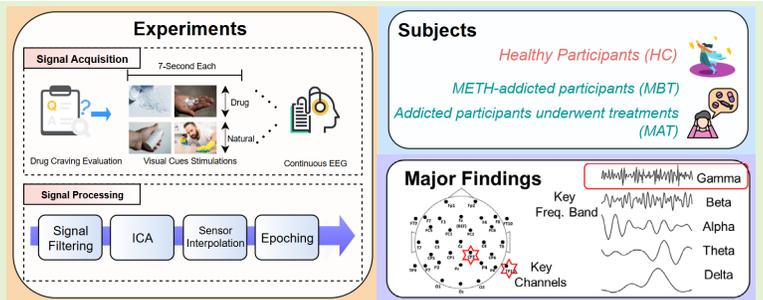

## I. Introduction

ADDICTION is defined as an overwhelming urge to use a particular substance or engage in a specific behavior, often leading to harmful consequences. Addiction to one such substance, methamphetamine (METH), is termed as methamphetamine use disorder or dependence (MUD); this has been listed as a serious public health concern [1]. METH is a highly addictive synthetic central nervous system stimulant. METH users experience positive feelings such as euphoria, increased self-confidence, and heightened energy levels in the short-term following use. MUD not only causes physiological and mental problems for individuals [2] but also accelerates biological aging and can lead to severe facial appearance changes [3]. There is currently no approved pharmacotherapy treatment available for MUD [4]; however, behavioral interventions have proved effective [5]. One common type of behavioral intervention for MUD is abstinence-based treatment in rehabilitation centers, but relapse rates among MUD individuals remain substantial. A study examining youth using ketamine and METH suggests that METH users are more prone to relapse than those in the ketamine group [6]. Encountering drug-related cues may trigger intense cravings for METH, even after a prolonged period of abstinence [7]. To address the issue of relapse in MUD individuals, specialized research is being conducted on METH addiction mechanisms [8] and neural activity during the abstinence period [9].

This study was supported by Westlake University, Zhejiang Key R&D Program (Grant No. 2021C03002) and "Pioneer" and "Leading Goose" R&D Program of Zhejiang (Grant No. 2024C03040). Z. Zeng and Y.-H. Chen contributed equally and are the co-first authors. (Corresponding authors: Y.-H. Chen, M. Sawan.)

Ziyi Zeng, Yun-Hsuan Chen, Xurong Gao, Wenyao Zheng, Jie Yang, Chengkai Wang and Mohamad Sawan are with CenBRAIN Neurotech Center of Excellence, School of Engineering, Westlake University, Hangzhou 310030, China (e-mail: chenyunxuan@westlake.edu.cn).

Ziyi Zeng is also with the School of Data Science, The Chinese University of Hong Kong (Shenzhen), Shenzhen 518172, China (e-mail: ziyizeng@link.cuhk.edu.cn).

Hemmings Wu and Zhoule Zhu are with Department of Neurosurgery, Second Affiliated Hospital, School of Medicine, Zhejiang University, Hangzhou 310009, China.

Lihua Zhong is with the Department of Education and Correction, Zhejiang Gongchen Compulsory Isolated Detoxification Center, Hangzhou 310011, China.

Weiwei Cheng is with Zhejiang Liangzhu Compulsory Isolated Detoxification Center, Hangzhou 311115, China.

Neuromodulation techniques, such as repetitive transcranial magnetic stimulation (rTMS), have the potential to treat addiction disorders [10]. rTMS is a non-invasive technique that involves the use of a magnetic field by energizing an electromagnetic coil to generate a current that stimulates cortical neurons and alters neural activity in targeted brain areas. Multiple studies have shown that rTMS applied to the dorsolateral prefrontal cortex (DLPFC) can significantly reduce METH cravings or improve cognitive function in METH users [11]. Additionally, the FDA has approved the use of TMS for treating severe depression and obsessive-compulsive disorder [12], indicating its promising potential for treating addiction.

Nowadays, physicians assess the scores of the questionnaires (e.g., Desire for Drug Questionnaire) before and after treatment to determine treatment outcomes. However, questionnaires are a subjective approach to evaluate effectiveness, and results can vary for the same participant depending on different scenarios. For example, mood and environment may influence an individual's response. Additionally, limited studies have been conducted to investigate the biological heterogeneity of METH addicts. Thus, using objective approaches to evaluate treatment outcomes is preferable. Moreover, quantified assessments can be used to establish subsequent treatment plans.

As addiction is a neurological disorder, brain signal activations in individuals with addiction can differ from those without. Electroencephalography (EEG) is a well-known technique for recording real-time brain electrical signals using electrodes placed on the scalp. It is widely used because of its high temporal resolution, low cost, non-invasiveness, and convenience. EEG biomarkers applied to identify MUD include event-related potential (ERP) [13, 14], microstates [15], functional connectivity [16, 17], and spectrum [18, 19]. Spectrum analysis is popular for biomarker mining because raw temporal signals can be easily transformed into spectral signals using the Fourier transform. These spectral signals can be subsequently divided into five bands based on frequency range, namely delta, theta, alpha, beta, and gamma. These band powers can serve as statistical features for a comprehensive analysis of the overall brain state. Previous studies have reported increased delta and theta band powers in the EEG of METH-dependent individuals during resting states with closed eyes [20]. One study also found a decrease in the alpha band power [21], while another study reported that the ratio of delta to alpha band power increases when participants suffering from METH-induced psychotic disorder close their eyes during resting state [22].

In addition to analyzing EEG band powers during the resting state, drug-related cues are also applied to investigate variations in band power in response to visual stimuli. Research has shown an increase in beta and gamma band power and a decrease in delta and alpha band power when METH-dependent individuals are exposed to METH-related virtual reality (VR) cues compared to healthy participants [23]. Another study reported an increase in gamma band power in METH-dependent participants when watching VR videos with METH cues [1]. Similarly, another study demonstrated changes in each frequency band power along over a 20-min drug-related VR video [24]. Most biomarkers used to identify METH dependency are based on the average band power of all EEG channels across the scalp. Only two studies have considered the average band power of EEG channels in various cortical sub-regions as well [1, 21]. While most biomarkers are identified when presenting METH-related cues, to the best of our knowledge, only one study has compared the band power when receiving METH-related and neutral cues [23]. These above-discussed biomarkers are defined to distinguish the participants with or without METH dependency and are also necessary to characterize the effectiveness of treatments for individuals with MUD. Li et al. showed that the gamma band power of METH dependents becomes similar to that of healthy participants after undergoing a VR counter-conditioning procedure [1]. Another study showed a decrease in the frontal theta/beta power ratio after intermittent theta-burst stimulation, a type of TMS [25]. Nevertheless, limited studies have evaluated the effectiveness of treatment by comparing the EEG spectrum between patients after treatment with that of the healthy group.

One limitation of previous studies is the use of statistical data analysis, such as ANOVA and t-test, to identify EEG spectrum biomarkers for METH dependence. This method requires the hypothesis of potential biomarkers based on specific band power within specific cortical regions. However, this approach may inadvertently neglect other biomarkers that are not well-established or have not been previously studied. Essentially, statistical analysis depends on hypotheses to interpret results, potentially leading to confirmation bias and limiting exploration of alternative explanations or unexpected findings. Machine learning (ML) has been increasingly utilized in various fields, including computer vision [26], natural language processing [27], and healthcare [28-30]. Classification in ML involves training models to automatically differentiate categories. One study utilized a classifier to distinguish between METH-dependent and healthy participants based on EEG and galvanic skin response (GSR) data [23]. Another study compares the performance of different ML classifiers on distinguishing the METH patients and HC before and after receiving METH-related VR videos [31]. One advantage of applying ML in classification tasks for distinguishing between the two participant groups is that the importance of EEG signals from each channel can be ranked. Understanding the dominant brain region in classifying addicted and healthy participants can help establish intervention protocols and design evaluation paradigms for treatment efficacy.

We designed an experiment to compare EEG signals before and after TMS treatment using a paradigm that included METH-related and neutral image cues. Biomarkers were determined by analyzing the EEG signal spectrum through both statistical data analysis and ML approaches. The remaining parts of this paper include in Section II the Materials and Methods showing the recruitment of participants and the experimental protocols. Also, dedicated signal processing and analysis procedures are presented in this section. Results and discussions are the subjects of Sections III and IV, respectively,

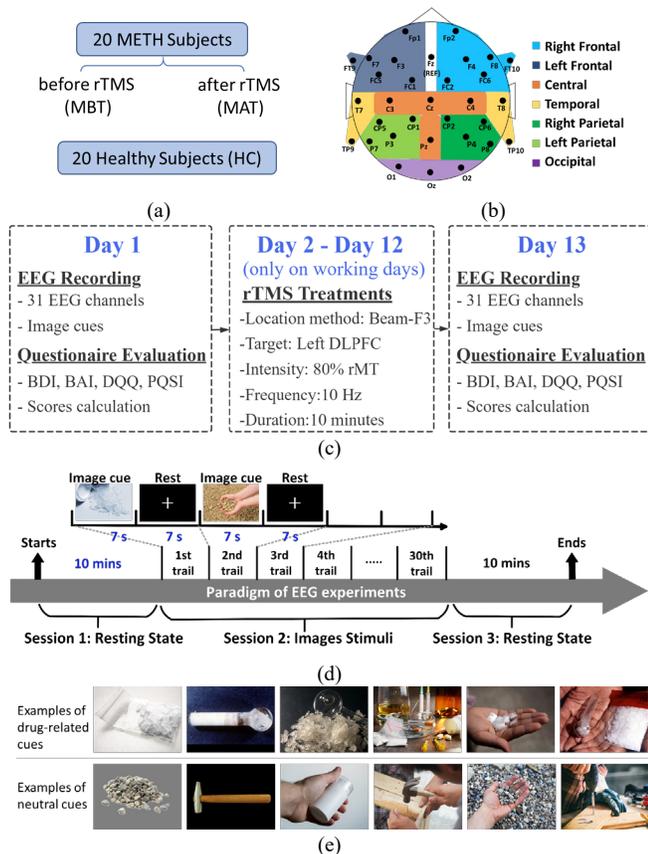

Fig. 1. Experimental protocols evaluating the EEG spectrum on METH-addicted participants before and after rTMS treatment and the control group: (a) Participant groups; (b) EEG channel locations and corresponding brain regions defined here; (c) Timeline of EEG measurements and TMS treatment; (d) Paradigm of EEG measurements; (e) Examples of METH-related (drug) and neutral cues. and the paper conclusions are in Section V.

## II. MATERIALS AND METHODS

### A. Participants

All participants with a history of METH dependence (METH group) were recruited from Gongchen Rehabilitation Center, Zhejiang, China. The criteria for participating in this study are listed in Table SI. The main recruitment criteria are as follows: 1) Age between 18 and 50 years; 2) Must meet the DSM V diagnosis of MA use disorder; 3) Must have had at least 4 weeks of detoxification and wish to stop using MA. Throughout the treatment period, the participants received no other treatment than rTMS. Twenty-four drug-dependent individuals were recruited, but four participants were excluded due to concurrent polydrug usage. Consequently, twenty participants' recording were retained for further analysis. The METH-addicted participants Before and After rTMS formed MBT and MAT, respectively. Meanwhile, participants who had never used METH were recruited from Westlake University by spreading the recruitment information through social communication groups. The criteria for participation are listed in Table SII. Thirty male volunteers were recruited as the healthy group (HC). After excluding recordings that were not completed due to technical problems and those with poor signals using the evaluation methods (Section II.D), a total of twenty recordings were retained. Fig. 1a shows the division of each group. Informed consent was obtained from all participants in both groups before their involvement in the study.

### B. Experimental Procedure

#### 1) Experimental protocol for METH participants

An EEG recording system (Brain Products, USA) equipped with an actiCHamp Plus amplifier and actiCAP slim active EEG electrodes was used. Thirty-three electrodes were mounted on the scalp according to the international 10-20 standard (Fig. 1b). The reference electrode was placed at Fz and the ground electrode was placed at Fpz forming a 31-channel recording. EEG signals were recorded at a sampling rate of 500 Hz, and no filter was applied during the recordings. On assessment days (Day 1 and Day 13), participants completed four questionnaires: Desire for Drug Questionnaire (DDQ), Beck Depression Inventory (BDI), Beck Anxiety Inventory (BAI), and the Pittsburgh Sleep Quality Index (PSQI). After completing the questionnaires, EEG measurements began. Participants were asked to sit on a comfortable chair with a screen in front displaying instructions and visual stimuli during the recordings. All subjects were asked to focus on the screen and avoid moving their bodies. During the study period, from Day 2 to Day 12, each participant received a 10-minute rTMS treatment daily for 10 days (Fig. 1c). The detailed TMS protocol is provided in Section II.C. The 10-day TMS treatment was not conducted consecutively due to the unavailability of researchers and physicians at the rehabilitation center on weekends.

The protocol for EEG measurements is shown in Fig. 1d. Session 1 involved a 10-minute resting period. The participants were asked to close their eyes but not fall asleep. Session 2 involved the presentation of image stimuli. Fifteen METH-related images and fifteen neutral images were chosen from the Methamphetamine and Opioid Cue Database (MOCD) [32]. The neutral images in the database were intentionally selected to have some degree of association with the drug-related images, such as matching content (objects, hands, faces, and actions). Neutral images were selected based on craving scores that ranged evenly from the highest to the lowest reported in the database. Examples of chosen images are shown in Fig. 1e. During Session 2, the METH-related and neutral images were presented randomly. Each image appeared on the screen for 7 seconds, followed by a prompt asking individuals to rate their craving level based on the image cue. A 7-second rest period followed each image presentation. During this rest period, participants were asked to focus on a "+" sign in the center of the screen to minimize head movement. Each cycle of presenting an image and rest period was considered a trial. After 30 trials, a 10-minute resting period preceded the conclusion of the measurement (Session 3). The protocol was approved by the ethical committees of Westlake University (ID: 20191023swan001) and The Second Affiliated Hospital Zhejiang University School of Medicine (ID: 2023_0522).

#### 2) Experimental protocol for healthy participants

The healthy group did not complete any of the four questionnaires and did not undergo TMS treatment. The healthy participants underwent the same EEG measurements as the METH participants. To study the change of brain signals of addicted participants is mainly resulted from the rTMS, meaning is not influenced by the repeated measurement protocol, five health subjects were invited to participate in a second phase measurement with the same protocol after 1 month (HCA).

### C. TMS Treatment

The rTMS protocol was conducted once per day, and each participant received 10 days of treatment following the procedure outlined by [33]. The mode was iTBS, with the parameters set at 80% of the active motor threshold, repeated at 10 Hz, with 5 s on and 10 s off, for a total duration of 10 minutes and 2000 pulses. The stimulation location was determined using the Beam-F3 method, with the round coil placed on the subject's left DLPFC at a point 5 cm anterior to the scalp position where the motor threshold was determined. The stimulation commenced by clicking the start button and stopped when the time had elapsed.

### D. EEG Signal Processing

1) *Signal pre-processing*

First, we excluded recordings if at least 80% of the channels were heavily contaminated by 50 Hz signals (power exceeding 8 dB after Fourier transform). The signals were then filtered using an IIR band-pass filter at 0.5 to 60 Hz, followed by a notch filter at 50 Hz to remove powerline interferences. Channels with high impedance exceeding 200 kΩ and those with visibly abnormal shapes upon visual inspection were identified as bad channels. The signals from these channels were replaced with surrounded signals via spherical spline interpolation. These procedures were conducted using BrainVision Analyzer (Brain Products GmbH, Gilching, Germany). Subsequently, independent component analysis (ICA) was employed to decompose the EEG data into a series of components using FastICA as defined in [34] and conducted in MNE-Python [35, 36]. During FastICA, artifact components including electrooculography (EOG) and electromyography (EMG) were selected for auto-rejection. EOG components were rejected if the z-score exceeded a certain threshold, while muscle components were rejected if the correlation with the typical muscle component surpassed a given threshold. Then, the signals were segmented into 7-second epochs during which METH-related and neutral images were displayed.

2) *Calculate relative band power (RBP)*

After signal preprocessing, the multi-taper method (MTM) was utilized to convert the EEG signals into power spectrum density (PSD). The PSD of the 7-second epochs was calculated for individual channels, each brain region, and all channels. Seven brain regions were identified for this study (Fig. 1b): left frontal, left parietal, occipital, central, temporal, right parietal, and right frontal lobes. The relative band power (RBP) of each channel was calculated using (1) to compare the EEG spectrum recorded from various channels across different participants.

$$RBP_{band} = \frac{1}{(b-a)} \sum_{i=a}^{b} RBP_i = \frac{1}{(b-a)} \sum_{i=a}^{b} \frac{ABP_i}{\sum_{j=1}^{44} ABP_j} \quad (1)$$

where $RBP_i$ represents the relative band power of a specific frequency band i. The $RBP_i$ was derived by the absolute band power (ABP) at frequency i divided by the summation of ABP of each frequency point in the EEG effective range (1-44 Hz). To calculate the RBP of the five frequency bands (delta: 1–4 Hz, theta: 4–8 Hz, alpha: 8–13 Hz, beta: 13–30 Hz, and gamma: 30–44 Hz), a was defined as the starting frequency and b is defined as the end frequency of each sub-band range, respectively.

The pre-processing and analysis code used for this study is available at: https://github.com/ZiyiTsang/Assess_EEG_Effectiveness_rTMS.

### E. Statistical Analysis of RBP

Each participant generated 15 epochs of drug-related cues and another 15 epochs of neutral cues. The RBP values of these 15 epochs were averaged when determining the RBP of a certain channel. The averaged RBP values of those channels from the same brain region were averaged to be defined as the RBP of a certain brain region. Then, the averaged RBP values of all 31 channels are averaged to be shown in Section III-B.

To analyze the EEG response of image cues over time, the epochs' length was redefined as 3.5 s in the last part of the experiment. The RBPs before the image cues, during the first and second phases of the cues, and after the cues were calculated.

### F. Statistical Analysis of the Questionnaires and RBP

SciPy (version 1.10.1) in Python was used for statistical analysis [37]. To measure treatment effectiveness, a questionnaire was used for statistical analysis. The Wilcoxon signed-rank test was used to compare the scores of the questionnaires before and after treatment because all four questionnaire scores did not follow a normal distribution. For other demographic information and RBP, independent sample t-tests were used for comparisons between the METH and control groups. The false discovery rate (FDR) correction was used in the t-tests to avoid bias from multiple tests. Compared to other methods (e.g., Bonferroni correction), FDR provides an optimal balance between type I and type II error, which is critical in our case with numerous correlated EEG channels and frequency bands analyzed concurrently. Furthermore, to ensure the robustness of our result, effect size calculations were performed using Cohen's d tests on gamma RBP across all brain regions between MBT and HC.

### G. ML Analysis of RBP

Each subject has 15 trials (epochs) of drug cues and 15 trials (epochs) of neutral cues. Those trails from the same type of cue and the same participant's group were concatenated together for the following ML analysis and classification.

Both Support Vector Machines (SVM) and the random forest (RF) algorithm were applied for classification. Although the performance of RF is similar to that of SVM, RF does not need to include complex hyperparameter tuning as SVM usually does. Most critically, RF's capacity for generating interpretable feature importance metrics (via MDI and SHAP) provides essential mechanistic insights into the neural substrates of methamphetamine dependence. Therefore, we only present the details of the RF model in this paper. This model employed 100 estimators, with the gini loss function. First, the classification based on each sub-band was carried out respectively. During each classification, the total number of features was 39 (31 RBP values of each channel, 7 RBP values of each brain region and 1 RBP value of all channels). Moreover, to better represent changes in overall EEG frequency sub-bands (1-44 Hz), each feature from the 5 sub-bands was averaged to perform an integration classification. In this way, the number of features was the same as the previous task, avoiding adverse effects on RF due to the different numbers of features. Previous studies have used feature selection (FS) methods such as recursive feature elimination as these have been shown to improve classification performance [38]. However, FS can cause feature imbalance within several classification tasks, making it difficult to make comparisons. Therefore, for a fairer comparison, FS was not used in our study. The 5-fold cross-validation, referred to in a paper using RBP of EEG for classification, was used to achieve robust results [39]. The F1-score served as an evaluation metric for validation. See (2). The model was implemented based on Scikit-learn (version 1.2.2) [40].

$$F1 = \frac{2PR}{P+R} \quad (2)$$

where P is the precision rate, and R is the recall rate of the result.

To rank features for classification, we utilized both Shapley Additive Explanations (SHAP toolbox with version 0.43) and Mean Decrease in Impurity (MDI). SHAP provides model-agnostic, instance-level feature importance by quantifying each feature's marginal contribution based on game theory. MDI, assessed via 5-fold cross-validation (Scikit-learn toolbox with version 1.2.2), measures importance based on the cumulative impurity reduction within RF. These methods collectively evaluated the prominence of 31 gamma-band EEG channels in differentiating MBT from HC.

## III. RESULTS

### A. Demography and Questionnaire Scores

The demographic information of the METH-addicted and healthy groups is shown in Table I. Number of detoxification times means the number of detoxification processes that were obliged in a rehabilitation center for each METH subject. There are differences in age and years of education between the two groups. However, as our experimental paradigm does not involve mental workload tasks or require quick reactions, which can be influenced by age and education, we believe our results can still yield meaningful conclusions. In terms of the questionnaires, the score of DDQ decreased significantly after rTMS treatment (p<0.01), as shown in Fig. S1. Additionally, a noteworthy decrease in the BAI score was observed (p<0.01). The mean value of the BDI indicator decreased from 6.4 to 4.1 (p=0.06). However, there was only a slight change in the PSQI indicator before and after TMS (p=0.49).

### B. Statistical Methods to Compare METH and Healthy Groups

RBPs are presented here when receiving METH-related and neutral cues. The statistical data analysis methods introduced in Section II to distinguish between the METH-addicted and healthy groups are also presented.

1) *Resting-state RBP*

The RBPs during resting states of two periods (Sessions 1 and 3 in Fig. 1d), before and after receiving cues, are plotted in Fig. 2. In both resting periods, the healthy group shows higher delta and alpha RBPs compared to the METH-addicted group. Meanwhile, HC has lower theta, beta, and gamma RBPs. When examining the effects of visual stimuli on RBP, it is observed that the delta RBP increases after receiving cues for all groups. Additionally, a decrease in theta, beta, and gamma RBP is observed across all groups. Both MAT and HC show a decrease in alpha RBP after the image cues. However, the alpha RBP of MBT does not exhibit a significant difference before and after the visual stimuli.

TABLE I
DEMOGRAPHIC INFORMATION OF METH-ADDICTED AND HEALTHY GROUPS. UNITS: YEARS OR TIMES ± STANDARD DEVIATION; *: THE P-VALUE OF 2 GROUPS <0.01.

| Demographic Information | METH (n=20) | Healthy (n=20) |
|---|---|---|
| Mean age* | 36.90 ± 7.72 | 26.10 ± 4.20 |
| Mean age at first METH use | 35.70 ± 8.20 | N/A |
| Number of detoxification times | 1.65 ± 0.99 | N/A |
| Mean years of education* | 8.30 ± 2.80 | 18.10 ± 1.50 |

2) *RBP when receiving METH-related cues*

The average RBP of all EEG channels at each frequency band is shown in Table II. The RBPs of MBT and MAT are compared at each frequency band. Following treatment, the theta and alpha band waves of the METH group show a significant increase. Conversely, there are decreases in fast waves without statistically significant differences. For the alpha, beta, and gamma bands, the RBPs of METH individuals become similar to those of the healthy group after treatment. To prove the variations are mostly due to rTMS, the RBP on the healthy group in the first and second measurements are derived, shown as HC and HCA in Table II. The changes of delta, alpha and gamma RBP between HCA and HC during drug-related cues can be neglected (<0.01). This shows that these RBP values do not vary much when healthy people participate in the same EEG protocol one month after the first measurement. This further proves that the larger alpha and gamma RBP values' change which brings the MAT's value closer to that of HC's value are due to TMS treatment. Regarding beta RBP values, although the change of MAT versus MBT is smaller than HCA versus HC, the variation trend still shows a "normalization" (MAT's beta RBP is closer to HC's value compared to MBT's).

To better visualize the RBPs of individual channels during METH-related cues, Fig. 3 shows the topographies of each frequency band. The average RBP of each cortical sub-region is plotted in Fig. 4a. Compared to HC, there is an increase in the theta, beta, and gamma RBP across all sub-regions for MBT. In terms of alpha RBP, although MBT shows increased value compared to MAT, the HC still exhibits significantly higher values in the parietal lobe compared with the addicted groups. Following TMS treatment, a reduction in beta and gamma RBP is observed in the parietal and occipital lobes of MBT, approaching levels like those of HC. Furthermore, the temporal lobe also notes a decrease in gamma RBP. Regarding the gamma RBP, the differences between MBT and MAT are significant (p<0.05) at the above-mentioned brain regions with Cohen's d values 1.08 at L_pareital, 1.11 at R_parietal and 0.98 at temporal regions. Therefore, in beta and gamma bands, the topographies for the METH population become more comparable to those of healthy individuals after TMS treatment. However, it is worth mentioning that there exists a large gap between MAT and HC in the theta, alpha, beta, and gamma bands.

To better illustrate RBP changes across conditions, Fig. 4b shows the relative power spectrum of 3 groups of participants at parietal and temporal regions. This presents a more detailed comparison in the frequency band. In the conventionally defined beta band (13-30Hz), we find that the RBP of 18-24 Hz at the parietal region can be more suitable to be used as a biomarker to distinguish the MBT and MAT.

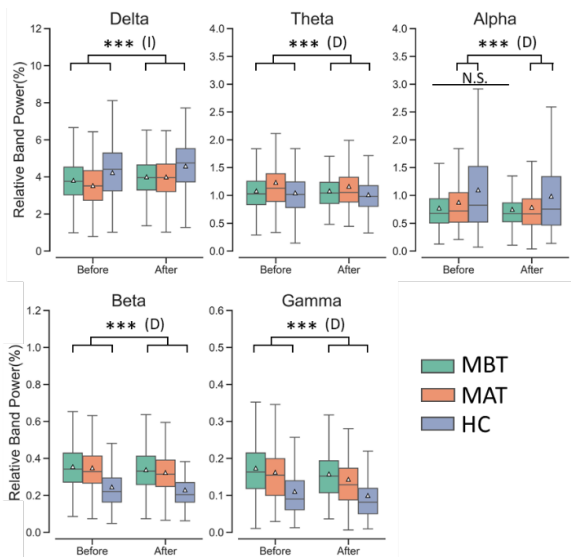

Fig. 2. Relative band power of each frequency band before and after picture stimuli in EEG recordings. ***: $p<0.01$. I: RBP of the resting state after stimuli is larger than that before stimuli. D: RBP of the resting state after stimuli is smaller than that before stimuli. N.S.: no significant difference.

TABLE II
ALL EEG CHANNELS' AVERAGE RBP AT EACH FREQUENCY BAND WHEN RECEIVING DRUG-RELATED CUES. SD ARE SHOWN IN PARENTHESES.

| | Bands | Delta | Theta | Alpha | Beta | Gamma |
|---|---|---|---|---|---|---|
| | MBT | 2.050 ± 0.426 | 0.602 ± 0.159 | 0.338 ± 0.114 | 0.163 ± 0.046 | 0.091 ± 0.036 |
| | MAT | 2.002 ± 0.549 | 0.644 ± 0.256 | 0.363 ± 0.155 | 0.159 ± 0.055 | 0.085 ± 0.054 |
| | HC | 2.300 ± 0.559 | 0.592 ± 0.156 | 0.445 ± 0.224 | 0.118 ± 0.048 | 0.056 ± 0.037 |
| | HCA | 2.310 ± 0.775 | 0.535 ± 0.110 | 0.435 ± 0.346 | 0.134 ± 0.056 | 0.054 ± 0.040 |
| P-value of | MBT & HC | <0.01 | 0.43 | <0.01 | <0.01 | <0.01 |
| | MBT & MAT | 0.19 | <0.01 | <0.01 | 0.24 | 0.09 |
| | MAT & HC | <0.01 | <0.01 | <0.01 | <0.01 | <0.01 |

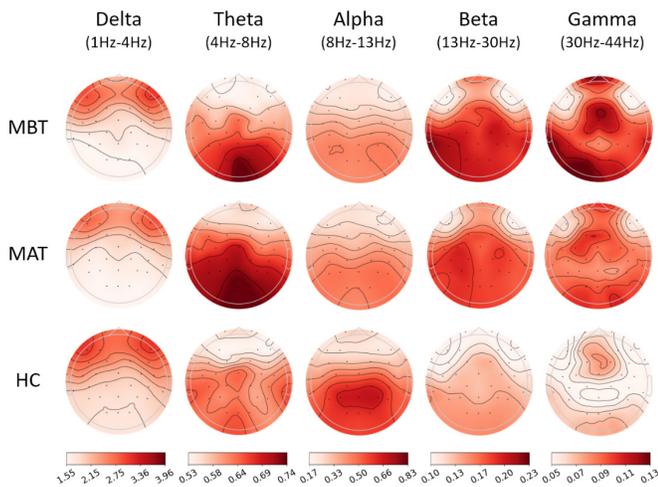

Fig. 3. Topographies of RBP at various frequency bands when receiving drug-related cues.

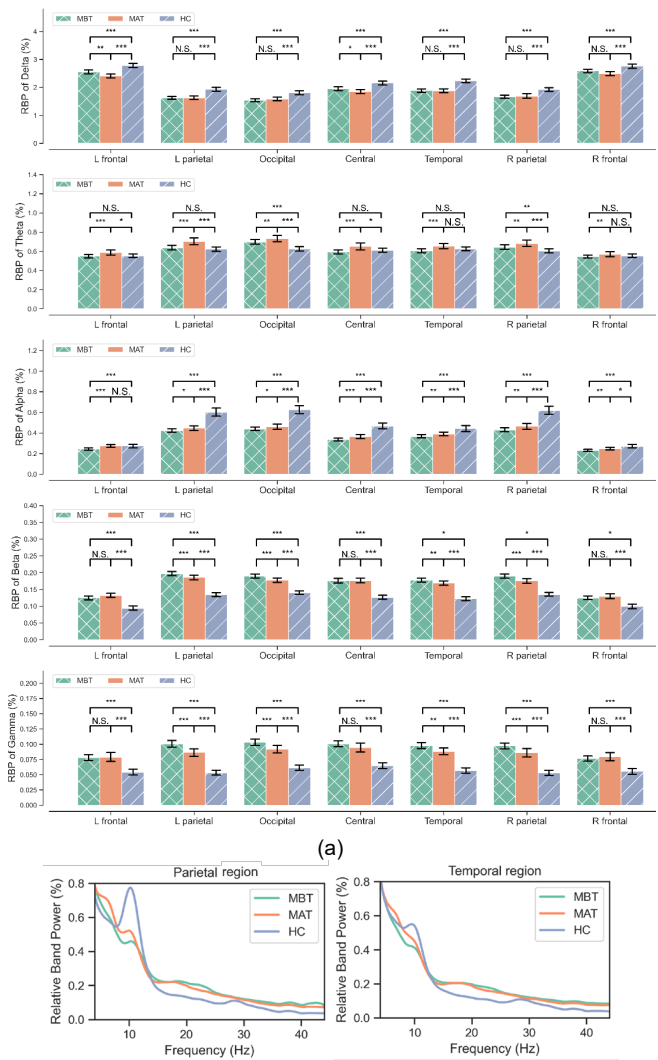

Fig. 4. (a) RBP at all sub frequency bands when receiving drug-related cues. ***: $p<0.01$; **: $0.01<p<0.05$; *: $0.05<p<0.1$; N.S.: $p>0.1$. (b) Relative power spectrum of the MBT, MAT and HC at parietal and temporal regions.

3) *RBP when receiving neutral cues*

The average RBPs of all EEG channels and each sub-region at different frequency bands are shown in Table SIII, and Fig. S2 and S3. The RBP of the alpha, beta, and gamma bands in the METH groups either increases or decreases towards the levels observed in the healthy group after TMS treatment. This trend, however, is not observed in the delta and theta bands. Similar to receiving the visual stimulation via METH-related cues, an increase in beta and gamma RBP is seen in MBT compared to HC across all subcortical regions. A slight increase in the theta band is only observed in the left frontal, occipital, and right parietal regions. Additionally, a decrease in alpha wave RBP is seen in MBT across all subregions. After TMS, the RBP of alpha, beta, and gamma becomes more similar to that of HC, particularly noticeable in the right and left parietal lobes and the occipital lobe.

### C. ML Analysis to Compare Between the Participants' Groups and Between the Visual Cues

Table III shows the ML results of classifying group pairs based on the RBP features of each frequency band or on an integrated feature across all bands using RF's F1 scores. AUROC and accuracy values are in Tables SIV and SV.

TABLE III
F1 SCORE AND STANDARD DEVIATION OF CLASSIFYING THREE PAIRS OF TWO GROUPS BASED ON VARIOUS RBPs OF EACH FREQUENCY BAND WHEN RECEIVING DRUG-RELATED AND NEUTRAL CUES. THE CLASSIFICATION WAS PERFORMED USING AN RF MODEL.

| | Group | Delta | Theta | Alpha | Beta | Gamma | Integration |
|---|---|---|---|---|---|---|---|
| Drug-cue | MBT v.s HC | 0.82 ± 0.03 | 0.75 ± 0.05 | 0.80 ± 0.03 | 0.88 ± 0.02 | 0.90 ± 0.02 | 0.87 ± 0.02 |
| | MAT v.s HC | 0.80 ± 0.02 | 0.71 ± 0.04 | 0.78 ± 0.04 | 0.82 ± 0.06 | 0.88 ± 0.01 | 0.83 ± 0.03 |
| | MBT v.s. MAT | 0.74 ± 0.03 | 0.62 ± 0.05 | 0.61 ± 0.03 | 0.77 ± 0.04 | 0.84 ± 0.03 | 0.77 ± 0.02 |
| Neutral-cue | MBT v.s HC | 0.74 ± 0.04 | 0.75 ± 0.03 | 0.74 ± 0.03 | 0.82 ± 0.04 | 0.88 ± 0.04 | 0.80 ± 0.06 |
| | MAT v.s HC | 0.73 ± 0.03 | 0.71 ± 0.02 | 0.77 ± 0.03 | 0.79 ± 0.04 | 0.87 ± 0.03 | 0.81 ± 0.04 |
| | MBT v.s. MAT | 0.71 ± 0.05 | 0.65 ± 0.03 | 0.60 ± 0.03 | 0.75 ± 0.03 | 0.80 ± 0.02 | 0.76 ± 0.03 |

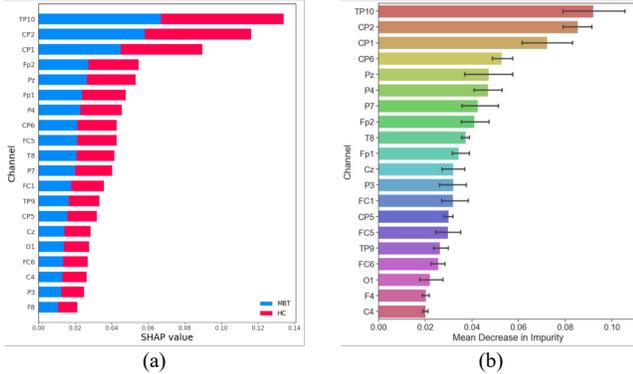

Fig. 5. The channel importance is determined by (a) SHAP and (b) MDI. ci=95%.

TABLE IV
F1 SCORE AND STANDARD DEVIATION OF CLASSIFYING RBP DURING METH-RELATED AND NEUTRAL CUES BASED ON EACH FREQUENCY BAND.

| | Delta | Theta | Alpha | Beta | Gamma | Integration |
|---|---|---|---|---|---|---|
| MBT | 0.50 ± 0.03 | 0.53 ± 0.06 | 0.49 ± 0.05 | 0.44 ± 0.05 | 0.48 ± 0.04 | 0.51 ± 0.06 |
| MAT | 0.53 ± 0.03 | 0.51 ± 0.03 | 0.52 ± 0.04 | 0.51 ± 0.03 | 0.49 ± 0.04 | 0.54 ± 0.04 |
| HC | 0.52 ± 0.05 | 0.49 ± 0.03 | 0.51 ± 0.04 | 0.55 ± 0.07 | 0.47 ± 0.03 | 0.50 ± 0.05 |

Although the advantages of RF were mentioned in Section II.G., the classification performance using SVM is presented in Table SVI for comparison. In Table III, the classification results of MBT versus HC were slightly higher than those of MAT versus HC across all bands, indicating greater similarity in EEG features between MAT and HC. Additionally, the classifier's performance during METH-related cues was superior to neutral cues both before and after TMS treatment. Among the frequency bands, the gamma band yielded the highest value in classification between MBT and MAT (F1=0.84). It is worth noting that the classification performance of the integrated feature was slightly lower than that of the gamma band, a trend observed for both METH-related and neutral cues.

In Fig. 5, the dominant channels to distinguish MBT and HC are ranked using features important analysis via SHAP and MDI. Regarding the SHAP analysis, TP10 channel was identified as the most critical channel, followed by CP2 and CP1. In MDI analysis, the same three channels dominate the RF classifier. This suggests that TP10, CP2 and CP1 play key roles in decision-making during RF classification for MBT and HC.

Moreover, the impact of drug-related and neutral cues on the EEG spectrum is investigated. Table IV presents the classification performance when distinguishing drug cues from neutral cues within each group of participants. The F1 scores across all groups are close to 0.5, suggesting random classification. This indicates that the classifier cannot accurately differentiate between drug-related cues and neutral

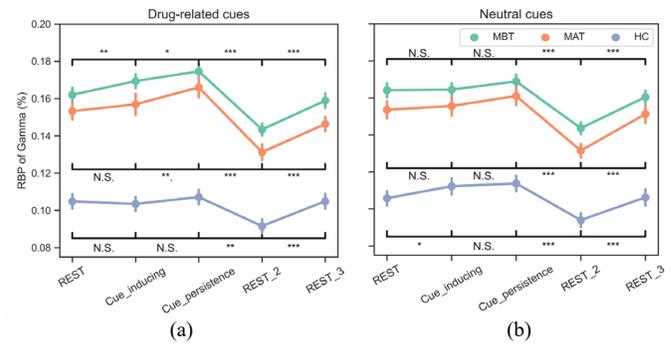

Fig. 6. Effect of cues with time. RBP of gamma frequency band of the three groups before, during, and after receiving: (a) drug-related, and (b) neutral cues. Error bar: stand error (se). ***: p<0.01; **: 0.01<p<0.05; *:0.05<p<0.1; N.S.: p>0.1.

cues based on the RBP derived from the 7-second epochs in which the cues are displayed.

### D. Effect of METH-related and Neutral Cues with Time

Given the unsatisfactory performance in classifying METH-related and neutral cues using the gamma RBP derived during the 7-second epochs, we attempt to optimize the epoch length for RBP calculation. Fig. 6 shows the RBPs at each 3.5 s time slot, spanning from 3.5 s before the image cues (REST) to 7 s after the cues disappear (REST_3). At the beginning of the REST_2 time slots, participants are prompted to rate the level of craving induced by the displayed image cues. Upon providing their response, the screen transitions to display a centered "+," signaling participants to rest. For METH participants, the period from the question appearing on the screen to pressing the keyboard averages at 0.95 s. When METH-related cues are presented, the gamma RBP for addicted participants keeps increasing until the craving level-related question appears, indicating a progression from REST to Cue_persistence. Conversely, the RBP for healthy participants does not increase significantly from REST to cue_inducing or to cue_persistence. Similarly, there is only a subtle increase in RBP among addicted participants during neutral cues from REST to cue_inducing (MBT: p=0.93, MAT: p=0.68) and from cue_inducing to cue_persistence (MBT: p=0.17, MAT: p=0.24). In contrast, the RBP for healthy participants increases from REST to cue_inducing and remains stable from cue_inducing to cue_persistence.

Regarding the impact of TMS, the growth of Gamma RBP from REST to cue-inducing became negligible in MAT (p=0.44) compared to that in MBT (p=0.04), mirroring the trend observed in HC (p=0.63). This suggests that the heightened gamma signals diminish following TMS treatment. In all cases, the gamma RBP decreases at REST_2 (when the craving level question is shown), followed by an increase at REST_3. The RBP values at REST_3 are similar to those at REST, indicating a return to baseline neural activity prior to the next cue. The RBP curves do not intersect across the three groups, with MBT leading, followed by MAT and HC.

### IV. DISCUSSION

We aim in this study to investigate the effect of rTMS treatment by assessing changes in EEG spectra from multiple aspects. In terms of the questionnaires, the DQQ, BAI, and BDI scores decreased after TMS treatment, which is consistent with the results reported in other studies [41, 42]. These findings suggest that our treatment not only reduces the subjects' cravings for drugs but also decreases anxiety levels and improves their sleep. Moreover, a strong correlation was observed between the changes in scores of DDQ and PSQI (r=0.71), aligning with the results of another study [43]. The similar impact on DQQ, BAI, and PSQI indicates that the TMS protocol used is effective.

To prove age and education differences between the METH and healthy groups do not have great impacts on our results, previous publications discussing this issue have been investigated. Regarding age, one paper shows that elder participants (ages 40-63) show lower alpha, delta and theta power than younger participants (ages 23-33) [44]. However, our results show that the elder group (HC with ages 36.90 ± 7.72) has higher alpha and delta power, which implies that the power difference does not result from the age gap. Regarding education level, one published paper found that men with higher education levels have higher relative gamma band power compared to those with lower education levels [45]. The authors claim that the higher gamma-band power of participants with higher education levels is due to higher cognitive load in their everyday life. In our study, the RBP of gamma of participants with lower education lever, meaning the METH-addicted participants (both MBT and MAT), is higher than that of the higher educated healthy group when receiving visual cues. This implies that education levels do not dominate the RBP of gamma values in our study.

In the resting state, we observed that the alpha RBP is smaller compared to HC, in line with previous research. One study demonstrated that the alpha power of all cortical regions in METH participants decreases compared to the healthy group when they are lying on a bed with their eyes closed [46]. The reduction in the alpha band is also evident when comparing METH participants and those with other drug use disorders to the healthy group [21]. No other studies have discussed the effect of receiving cues on resting state RBP. We found that the resting state alpha RBP decreases after receiving image cue stimulation in MAT and HC, but this was not observed in MBT. Alpha waves appear when a person is awake but in a resting state, usually with their eyes closed. A higher alpha power during resting indicates a more relaxed state. The decrease in RBP of the alpha band may be due to the memory of the drug and neutral cues, preventing participants from fully relaxing after the visual stimulation. Therefore, changes in alpha RBP during different resting periods may serve as a potential biomarker to distinguish between METH addicted participants and healthy individuals, as well as validate the impact of rTMS.

As for during cue-stimulation, our result shows alpha, beta, and gamma RBP of METH individual received rTMS became more similar to that of the health group both in topographies and bar diagrams. Moreover, our statistical analysis results support that the alpha, beta, and gamma RBP of all six cortical subregions or of the overall channels can serve as biomarkers to distinguish between participants with METH addiction and healthy individuals. One study shows that gamma power increases at global, anterior, central, and posterior scalp regions when METH participants are exposed to METH-related VR videos [1]. It is worth noting that the gamma activities in the anterior and central regions are more pronounced than in the posterior region. In contrast, our results indicate that not only the beta and gamma power of the frontal lobes increase but also those of the parietal and occipital lobes. Moreover, our findings demonstrate that gamma power increases most significantly in the posterior region. This may be attributed to the use of image cues rather than VR video cues. The variation in the beta band is also noteworthy when studying addiction. One study reported an increase in the average beta and gamma band power of 5 channels (Fpz, AF7, AF8, TP9, and TP10) when METH participants viewed VR videos featuring a METH-related environment [23]. Another study found that cue-induced cravings peaked at 3 months after METH participants began abstinence, with the highest beta band power recorded during this period [9]. In our study, the average duration of stay for METH participants undergoing abstinence treatment in a rehabilitation center was 9.65 months (SD: 6.64). This suggests that the increase in beta power we observed may not be the highest due to a longer period of abstinence than 3 months.

Research has shown that gamma band activity is elevated during recognition tasks in both humans and rats, and deficits in memory following repeated METH exposure may be attributed to altered gamma band activity [41]. Gamma power is associated with cognitive processing and visual binding [42]. The attentional capture of salient cues heightens perceptual processing, engaging glutamatergic pathways that are central to the brain's reward system [47]. Simultaneously, cue-induced memories activate neural circuits associated with drug use experiences, further amplifying gamma synchrony. Elevated glutamate levels, which correlate strongly with gamma power, reflect this state of hyperarousal and craving. Consequently, when METH addicted participants are exposed to drug-related image cues, cravings may be triggered by visual stimuli. Conversely, healthy participants do not experience cravings in response to cues.

Therefore, our results suggest that individual gamma RBP in specific brain subregions can serve as a biomarker to distinguish between MBT and HC. However, to enhance user convenience, reducing the number of EEG channels while maintaining detection accuracy is a future trend. In this study, we ranked the importance of each EEG channel during the classification task of MBT versus HC. EEG signals recorded from TP10 dominated, indicating that the temporal lobe has the strongest association with drug cravings. Patel et al. suggest that the temporal lobes are involved in situational memory and emotion regulation [43]. Individuals exposed to drug-related cues may exhibit heightened memory responses and mood changes. Another study supports this idea by reporting abnormal functionality of the temporal lobes of chronic drug abusers [44]. These findings highlight the differential response of the temporal lobes to stimuli, distinguishing them from those of healthy individuals. Furthermore, signals from the left and right parietal lobes play significant roles in addiction identification in our study. In the future, focusing solely on EEG signals from the temporal and parietal lobes and analyzing gamma RBP may be sufficient to distinguish between METH addicts and healthy groups.

Accordingly, biomarkers to assess treatment effectiveness can be divided into global biomarkers and those specific to cortex regions. From the whole-brain perspective, the RBP of beta and gamma waves decreases after TMS treatment, with RBP values approaching those of the healthy group, consistent with the findings of [45]. In our study, we observed a decrease in gamma RBP in the parietal, occipital, and temporal regions after TMS treatment. For frontal cortices, although there was no significant change in the average gamma RBP after TMS, a decrease could be observed at Fp1 and Fp2 when evaluating individual channels at the prefrontal region (Fig. 3). TMS is thought to normalize this aberrant neuroactivity by modulating cortical excitability and restoring balance in the dopaminergic and glutamatergic systems [48]. This finding supports the potential of TMS for alleviating addiction-related neural disturbances. Li et al. reported an augmentation of prefrontal gamma oscillatory in METH participants when exposed to drug cues, with the gamma power diminishing after treatment and aligning with levels observed in healthy controls [1]. Wen et al. demonstrated a reduction in gamma power in METH users watching METH-related VR videos after receiving an intermittent theta-burst stimulation treatment [25]. Our findings are consistent with these previous studies.

TABLE V
COMPARISON OF STUDIES USING VARIOUS EXPERIMENTAL PARADIGMS AND ANALYSIS TECHNOLOGIES FOR METH ADDICTION DETECTION. BOTH TYPES: BOTH METH-RELATED AND NEUTRAL CUES, EC: EYES CLOSED.

| Ref. | Cues (Both types) | ML analysis (Feature ranking) | Main biomarkers (MBT v.s HC by default, MBT v.s. MAT for study with treatments [§]) |
|---|---|---|---|
| [1] | VR videos (×) | × | Increase: gamma power Decrease: gamma power [§] |
| [2] | Images (√) | × | Increases: P300 peaks |
| [9] | Videos (×) | × | Increases: beta power Decreases: theta and alpha power |
| [14] | Images (√) | × | Increases: P3-related late positive potential Decrease: P3 component |
| [17] | EC resting | √ (×) | Increase: beta connectivity [§] |
| [23] | VR videos (√) | √ (×) | Increases: beta and gamma power Decrease: delta, theta, and alpha power |
| This work | Images (√) | √ (√) | Increase: delta and alpha power Decrease: beta and gamma power |
|  |  |  | Increase: theta and alpha power [§] |

The variation trends in RBP in individual frequency bands in METH participants exposed to METH-related and neutral cues are similar due to the selection of neutral cues. Both cues were selected from the MOCD with neutral pictures still having some links with METH, in terms of shape of particle, tool, or action [29]. For METH participants who experienced cravings when seeing the neutral pictures, the neural reactions could be similar to when seeing a METH-related picture. This may explain why ML analysis did not classify the two types of cues satisfactorily. However, when the epoch length for RBP analysis was reduced, the differences in gamma RBP between the two cue types became more pronounced. Notably, METH participants exhibited a marked increase in gamma RBP upon initial exposure to drug-related stimuli (Cue-inducing), with this response persisting in subsequent phases (Cue-persistence). In contrast, neutral cue led to only a slight increase, which was not statistically significant. Therefore, comparing the rate of increase in gamma RBP in a shorter time slot (i.e., 3.5 s) in the MBT group could serve as a promising biomarker to differentiate between responses to METH-related and neutral cues. In addition, the RBP variation from REST to Cue_inducing phases becomes insignificant after METH-addicted participants received rTMS, proving our TMS protocol's effectiveness.

In terms of the performance of the ML algorithm to distinguish healthy individuals and substance abusers based on EEG spectra, our algorithm demonstrated a noteworthy performance, achieving an F1 score of 90%. This result slightly surpasses that of a study with similar experimental frameworks (F1 score of 88.62%) [23]. This suggests that our RF algorithm could serve as an optimal model for the classification of MBT versus HC using our EEG recording paradigm. However, it is important to acknowledge that disparities in data processing methodologies across different studies may introduce biases in classification results. Moreover, the lack of comparable works emphasizes the need for further exploration of ML techniques in this field.

In terms of channel importance between MBT and HC, our result shows the channels from temporal lobe (TP10), parietal region (CP2 and CP1) exhibited notable prominence. This is consistent with our statistical analysis. Our study pioneers the integration of ML-based channel importance ranking into substance abuse research, warranting further exploration in future investigations.

In our experiment, we utilized the RBP metric, which is derived from the normalization of ABP. The advantage of RBP is that it equalizes inter-channel comparisons on a standardized baseline. Additionally, RBP made our classifier training more robust in the ML analysis because of the normalized features used, suggesting that incorporating RBP may lead to improved accuracy in ML classification. When comparing our RF model, which includes gamma RBP features, with Ding et al.'s work, our model achieves higher accuracy than their RF model, which includes ABP of the five sub-bands and GSR [23].

Compared to other published works on EEG monitoring in METH addicted groups, our study presents a comprehensive analysis using both statistical and ML methods to investigate the impact of METH-related and neutral cues on METH-dependent participants before and after TMS (Table V). This approach allows for a more robust conclusion to distinguish between METH and healthy groups, as well as assess the efficacy of TMS treatments. Moreover, we evaluated feature importance during the classification task to identify the dominant EEG channels. This method can guide a more user-friendly and analytically effective approach to identify individuals with METH addiction, as the gamma RBP of either the temporal or parietal lobes alone is sufficient to draw conclusions.

Our results support real-time gamma RBP monitoring as a biomarker for closed-loop neuromodulation in METH addiction treatment. Patients with elevated baseline gamma RBP may benefit from more frequent TMS. For post-rehabilitation monitoring, detecting rapid gamma surges (e.g., via channels like TP10 and CP2) could trigger on-demand rTMS during cravings. It is noteworthy that implementing such systems for clinical use still faces challenges like real-time signal processing and miniaturized hardware.

However, there are a few limitations of this study. First, only male participants were recruited, despite extensive research on gender differences in drug reinstatement and dependency [49, 50]. Given that rehabilitation centers worldwide prefer single-gender programs, most studies related to METH abstinence treatment often focus on single genders [11]. Future studies should validate our proposed biomarkers on female METH users to provide a more comprehensive understanding. By comparing the potential biomarkers to distinguish between the before and after rTMS-treated male and female groups, various assessment approaches can be applied to evaluate the efficacy of rTMS treatments in different genders. Second, there were age differences between healthy controls and METH participants. Although we proved in Fig. S4 of supplementary materials that age shows limited impact on EEG results in our experiments, future research could explore the role of gender and age as biological variables in relation to addiction, abstinence, and relapse levels. Third, the small sample size in both groups is also a limitation. Increasing the number of participants would improve the stability and accuracy of the ML classification model. To further investigate neural activities related to craving, abstinence experiences, and the risk of relapse, the use of multimodal wearable neuroimaging techniques such as EEG-fNIRS could provide higher spatial resolution of neural signals.

## V. Conclusion

We used both statistical and ML analyses on EEG spectra to investigate potential biomarkers for distinguishing METH-dependent individuals from healthy participants, as well as for classifying METH-dependent individuals before and after TMS treatment. Statistically, during exposure to METH-related cues, the alpha RBP of MBT at all individual brain subregions was smaller than that of HC, while the beta and gamma RBP of MBT was larger than that of HC. After TMS treatment, the values of alpha, beta, and gamma RBP all became similar to those of HC. When using a RF model to group MBT and HC, the gamma RBP showed promise as a distinguishing factor. TP10 and CP2 channels played leading roles when ranking the dominance of EEG signals during RF. Additionally, we demonstrated that analyzing the rate of increase in gamma RBP during a 3.5 second epoch could determine whether a METH-related or neutral cue was presented to a participant with MUD. Furthermore, we observed that the alpha RBP during the resting state decreased for MAT and HC after receiving a series of image cues, while the changes in MBT were insignificant. These biomarkers can be utilized in closed-loop neuromodulation systems for treating METH addiction and improving treatment efficacy.

## Supplementary

The supplementary material file contains three tables and four figures to support the content of this paper.

## Acknowledgment

The authors acknowledge the participants who volunteered in this study as well as the staff in both Zhejiang Gongchen Compulsory Isolated Detoxification Center and Zhejiang Liangzhu Compulsory Isolated Detoxification Center (Director of the Detoxification Medical Center: Hua Shen) for arranging the logistics of the experiments. The authors also thank Shuaishuai Li, Xiaobo Ye, and Jichao Lei for assisting in conducting the TMS treatment in a rehabilitation center. The help provided by Tianjun Wang, Yankun Xu, and Yingjue Bian regarding giving suggestions in data acquisition and signal processing is highly appreciated as well. Moreover, the supports from the Integrated-on-Chips Brain-Computer Interfaces Zhejiang Engineering Research Center.

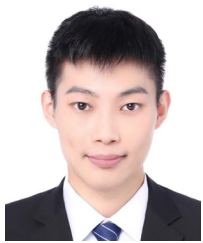

**Ziyi Zeng** received his B.Eng degree from Xiamen University, Xiamen, China, in 2024. He is currently pursuing the Msc degree with the School of Data Science, The Chinese University of Hong Kong-Shenzhen (CUHKSZ), China.

He holds a background in neuroscience, and artificial intelligence (AI), which bring multidisciplinary approach into his research. His current interests include the application of AI in medicine, especially in the medical large language model (M-LLM).

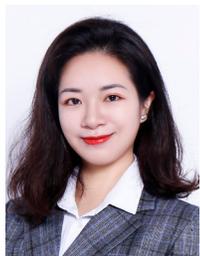

**Yun-Hsuan Chen** got her bachelor's degree in the Department of Materials Science and Engineering of National Tsing Hua University, Hsinchu, Taiwan. She got an Erasmus Mundus scholarship to pursue her joint master's degree in molecular nano- and bio-photonics for telecommunications and biotechnologies (Monabiphot) program in ENS Cachan, Paris, France, in collaboration with Complutense University of Madrid, Spain, and Delft University of Technology, the Netherlands. She received her Ph.D. in Electrical Engineering from the University of Leuven, Belgium, in 2016.

Then, she completed a postdoctoral fellowship in the Center of Excellence in Biomedical Research on Advanced Integrated-on-chips Neurotechnologies (CenBRAIN Neurotech) in 2021. Presently, Dr. Chen is a Research Assistant Professor in CenBRAIN Neurotech, leading research activities related to multimodal neuroimaging techniques for brain disorders.

**Xurong Gao** received the B.Eng. degree in electrical science and technology from South China Agricultural University, Guangzhou, China, in 2023. He is currently a Ph.D. candidate with the Center of Excellence in Biomedical Research on Advanced Integrated-on-chips Neurotechnologies (CenBRAIN Neurotech), Westlake University, Hangzhou, China, majoring in electrical science and technology. His research interests include decoding of visual neural signals and investigation of neural signal characteristics.

**Wenyao Zheng** received the B.S. degree in computer science, human-computer interaction from the University of Manchester, Manchester, UK, in 2020, and the M.S. degree in statistics from the National University of Singapore, Singapore, in 2021. She is currently working toward the Ph.D. degree in psychology from the University of St Andrews, St Andrews, UK. She was a Research Assistant with CenBRAIN Neurotech, Westlake University, China, from 2021 to 2024. Her research interests include mobile electroencephalography (EEG) and real-world cognitive neuroscience, with a focus on attention resource allocation and episodic memory.

**Hemmings Wu** received the M.D. degree in medicine from Peking University Health Science Center, Beijing, China, in 2006, and the Ph.D. degree in neuroscience from KU Leuven, Leuven, Belgium, in 2014. His major field of study was neurosurgery and neuroscience.

He is currently an attending neurosurgeon with the Second Affiliated Hospital, Zhejiang University School of Medicine, Hangzhou, China. Previously, he held research and clinical positions at Stanford University, KU Leuven, and other institutions. He has published extensively on neuromodulation, deep brain stimulation, and brain–computer interfaces. His current research interests include radiosurgical neuromodulation, the treatment mechanisms of brain stimulation for psychiatric disorders, and implantable brain–computer interfaces. Dr. Wu is a Board Member of the World Society for Stereotactic and Functional Neurosurgery. He has contributed to international guidelines on neurosurgery for psychiatric disorders. He received the Riechert Award in 2017.

**Zhoule Zhu** received his Ph.D. degree in neurosurgery from Zhejiang University, Hangzhou, Zhejiang, China, in 2023. He currently works as an Attending Physician in the Department of Neurosurgery, the Second Affiliated Hospital of Zhejiang University School of Medicine, Hangzhou. His current and previous research interests focus on the basic and clinical research of neuromodulation in neuropsychiatric diseases.

**Jie Yang** (Senior Member, IEEE) received the B.S. degree in electronic science and technology from Tianjin University, Tianjin, China, in 2010, and the Ph.D. degree in microelectronics from the Institute of Semiconductors, Chinese Academy of Sciences, Beijing, China, in 2015. From 2015 to 2019, he was a Postdoctoral Fellow with the I2Sense Laboratory, Department of Electrical and Computer Engineering, University of Calgary, AB, Canada. Since 2019, he has been with the School of Engineering, Westlake University, Hangzhou, China, where he is currently a Research Professor.

His research interests include circuits and systems for intelligent biomedical applications, mixed-signal system-on-chip design for brain–machine interfaces and neural decoding, and energy-efficient neuromorphic algorithms and architectures.

**Chengkai Wang** received the B.A. degree in information management and information systems from Hangzhou Dianzi University, Hangzhou, China, in 2025. He was a Visiting Student in the CenBRAIN Neurotech at Westlake University, Hangzhou, China, where his work focused on computational neuroscience and cognitive science research. He also serves as a Reviewer for several top-tier journals and conferences, including the IEEE Journal of Biomedical and Health Informatics (JBHI), NeurIPS, AISTATS and ICLR.

**Lihua Zhong** graduated from Wenzhou Medical College with a degree in Clinical Medicine in 2002, and from Zhejiang University with a degree in Clinical Medicine in 2009. Since 2002, he has been engaged in drug rehabilitation medical work at the Gongchen Mandatory Isolation Drug Rehabilitation Center in Zhejiang Province, serving as an associate chief physician.

**Weiwei Zheng** graduated from Zhejiang Judicial Police Vocational College (major in Prison Administration), obtained a correspondence undergraduate degree from Zhejiang University (major in Law), and holds an in-service graduate degree from the Provincial Party School (major in Philosophy). She has successively worked at the Zhejiang Provincial Drug Rehabilitation and Labor Camp, the Zhejiang Provincial Management Institute for Juvenile Labor Education and Rehabilitation, and the Zhejiang Provincial Liangzhu Compulsory Isolation Drug Rehabilitation Center. She has served as the director of the Education and Correction Center and the chief of the Drug Rehabilitation and Correction Department. She is skilled in education work and holds qualifications as a psychological counselor and a teacher.

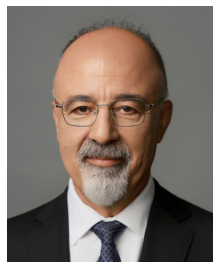

**Mohamad Sawan** (Life Fellow, IEEE) received the Ph.D. degree from the University of Sherbrooke, Sherbrooke city, Quebec, Canada. He is currently a Chair Professor with Westlake University, Hangzhou, China, and an Emeritus Professor with the Polytechnique Montreal, Montreal, Quebec, Canada. He is the Founder and the Director of the Center of Excellence in Biomedical Research on Advances-on-Chips Neurotechnologies (CenBRAIN Neurotech), Westlake University, Hangzhou, China, and of the Polystim Neurotech Lab, Polytechnique Montreal. He served as a Distinguished Lecturer of both IEEE CASS and IEEE SSCS for three consecutive years each. He was a Co-Founder, an Associate Editor, and an Editor-in-Chief of IEEE Transactions on Biomedical Circuits and Systems (2016–2019).

He was awarded the Canada Research Chair in Smart Medical Devices (2001–2015) and was leading the Microsystems Strategic Alliance of Quebec (ReSMiQ), Canada (1999–2018). He published more than 1000 peer-reviewed journal and conference papers, one handbook, three books, 13 book chapters, 15 patents, and 25 other patents are pending. He received several awards, among them the Barbara Turnbull Award from the Canadian Institutes of Health Research (CIHR), the J.A. Bombardier and Jacques-Rousseau Awards from the Canadian ACFAS, the Queen Elizabeth II Golden Jubilee Medal, the Medal of Merit from the President of Lebanon, the Chinese National Friendship Award, and the Shanghai International Collaboration Award. He is a Fellow of the Royal Society of Sciences of Canada (FRSC), a Fellow of the Canadian Academy of Engineering (FCAE), a Fellow of the Engineering Institutes of Canada (FEIC), and an "Officer" of the National Order of Quebec.